\documentclass[11pt]{article}
\usepackage[margin=1in]{geometry}
\usepackage[nosort]{cite}
\csname @addtoreset\endcsname{equation}{section}

\newcommand{\fft}[2]{{\frac{#1}{#2}}}
\newcommand{\ft}[2]{{\textstyle{\frac{#1}{#2}}}}
\newcommand{\sqr}[2]{{\vcenter{\vbox{\hrule height.#2pt
        \hbox{\vrule width.#2pt height#1pt \kern#1pt
        \vrule width.#2pt}\hrule height.#2pt}}}}

\newcommand{\be}{\begin{equation}}
\newcommand{\ee}{\end{equation}}
\newcommand{\bea}{\begin{eqnarray}}
\newcommand{\eea}{\end{eqnarray}}
\newcommand{\text}[1]{{#1}}
\begin{document}

%%%%%%%%%%%%%%%%%%%%%%%%%%%%%%%%%%%%%%%%

\begin{titlepage}
\hbox to \hsize{{\tt hep-th/0611244}\hss
\vtop{\hbox{MCTP-06-31}
\hbox{NSF-KITP-06-116}}}

\vspace*{1.5cm}

\begin{center}

{\bf \Large
All 1/2 BPS solutions of IIB supergravity with\\ $\mathrm{SO}(4)
\times\mathrm{SO}(4)$ isometry}

\vspace*{1cm}

James T. Liu$^{*,\dagger}$ and Wafic Sabra$^\ddagger$

\vspace*{.5cm}

{\it $^*$Michigan Center for Theoretical Physics,
Randall Laboratory of Physics,\\
The University of Michigan, Ann Arbor, MI 48109--1040, USA}
\vspace*{.4cm}

{\it $^\dagger$Kavli Institute for Theoretical Physics,\\
University of California, Santa Barbara, CA 93106, USA}
\vspace*{.4cm}

{\it $^\ddagger$Centre for Advanced Mathematical Sciences and Physics
Department,\\
American University of Beirut, Lebanon}

\end{center}

\vspace*{1cm}

\begin{abstract}
In hep-th/0409174, Lin, Lunin and Maldacena constructed a set of
regular 1/2 BPS geometries in IIB theory.  These remarkable `bubbling
AdS' geometries have a natural interpretation as duals of chiral
primary operators with weight $\Delta=J$ in $\mathcal N=4$ super-Yang
Mills theory.  Although these geometries have been assumed to be
complete, from a purely supergravity point of view, additional
1/2 BPS configurations may potentially exist with a preferred
null isometry.  We explore this possibility and prove that the only
additional class of 1/2 BPS solutions with $\mathrm{SO}(4)
\times\mathrm{SO}(4)$ isometry are the familiar IIB pp-waves.
\end{abstract}

\end{titlepage}

%%%%%%%%%%%%%%%%%%%%%%%%%%%%%%%%%%%%%%%%
\section{Introduction}

Motivated by the AdS/CFT correspondence, Lin, Lunin and Maldacena (LLM)
\cite{Lin:2004nb} recently explored the space of 1/2 BPS supergravity
solutions dual to chiral primary operators with weight $\Delta=J$
in $\mathcal N=4$ super-Yang Mills theory.  These solutions may be
viewed as 1/2 BPS excitations of AdS$_5\times S^5$, and are described
in terms of a `droplet' picture.  In particular, all such regular
1/2 BPS supergravity configurations are completely determined in terms
of boundary data (droplets) specified on a two dimensional ($x_1$--$x_2$)
plane.  As befits the AdS/CFT correspondence, this two dimensional
plane may be identified with the phase space of free
fermions that arise in the dual description of the corresponding
1/2 BPS states in the $\mathcal N=4$ gauge theory
\cite{Berenstein:2004kk}.

Although the 1/2 BPS geometries of IIB theory which are constructed
in \cite{Lin:2004nb} form a complete set of duals to the $\Delta=J$
chiral primaries, they do not necessarily comprise {\it all} 1/2 BPS
solutions of IIB supergravity.  For example, the basic supergravity
D3-brane itself is perfectly regular, and smoothly interpolates
between AdS$_5\times S^5$ and 10-dimensional Minkowski space.  Of
course, this full D3-brane geometry does not admit an
$\mathrm{SO}(4)\times \mathrm{SO}(4)$ isometry that
is relevant to $\Delta=J$ states, and hence does not fall in the
class of `Bubbling AdS' solutions considered in \cite{Lin:2004nb}.
Nevertheless, it is still an open possibility whether there are any
additional solutions with $\mathrm{SO}(4)\times \mathrm{SO}(4)$ isometry
that have not yet been identified.

The possibility that there are additional regular 1/2 BPS solutions of IIB
supergravity arises from two observations.  The first observation is that
the analysis of \cite{Lin:2004nb} only takes into account the metric and
self-dual five-form.  In principle, one could expect new solutions to arise
that involve the three-forms as well as the IIB dilaton-axion field.  It
was shown in \cite{Liu:2004ru}, however, that excitations of the
dilaton-axion generically lead to 1/4 BPS rather than 1/2 BPS solutions.
Although we have not investigated the possibility of turning on a
three-form, their non-trivial presence would almost certainly also lead to
further reduction of supersymmetry from 1/2 BPS to 1/4 BPS.  Thus we
believe that the space of 1/2 BPS configurations will not be enlarged
by turning on any additional fields.

However there is still the second observation, which is that the
construction of \cite{Lin:2004nb} using the invariant tensor approach of
\cite{Tod:1983pm,Gauntlett:2002sc,Gauntlett:2002nw,Gutowski:2003rg,Gauntlett:2004zh}
only follows the case where the Killing vector $K^\mu=\overline\epsilon
\gamma^\mu\epsilon$ formed out of the Killing spinor $\epsilon$ is timelike.
Here we recall that the classification and construction of such geometries
generally fall into two classes, depending on whether the Killing vector is
timelike or null.  In the case of bubbling AdS geometries \cite{Lin:2004nb},
the Killing vector $K^\mu$ is conjugate to the AdS energy (corresponding
to conformal dimension $\Delta$), so on physical grounds it ought to be
timelike.  However, as far as constructing {\it all} 1/2 BPS solutions is
concerned, one ought to investigate the null case as well for completeness.
In particular, we note that there are static solutions such as the magnetic
string in five dimensions which fall in the null Killing vector category
\cite{Gauntlett:2002nw}.

In this letter we complete the search for all 1/2 BPS solutions of IIB
supergravity preserving $\mathrm{SO}(4)\times\mathrm{SO}(4)$ isometry
by directly analyzing the null Killing vector case of the invariant
tensor construction.  The result is perhaps not so surprising: the
only solutions falling in this case are the IIB pp-waves with
appropriate isometry.  We note that the particular case of the maximally
supersymmetric pp-wave \cite{Blau:2001ne} is also be obtained in the timelike
case when the phase space configuration is identified with the half
filled plane.  Although our initial expectation, based on the analysis
of theories with eight supercharges \cite{Tod:1983pm,Gauntlett:2002nw},
was that the null case could capture a larger set of either pp-wave
or plane-fronted wave solutions, we found that this in fact does not
happen for IIB solutions with $\mathrm{SO}(4)\times\mathrm{SO}(4)$
isometry.  Nevertheless, our analysis essentially completes the
proof that all 1/2 BPS solutions of IIB supergravity with the given
isometry are of the
bubbling AdS form \cite{Lin:2004nb} (at least up to the remote
possibility of turning on three-form flux without breaking any
additional supersymmetry).

%%%%%%%%%%%%%%%%%%%%%%%%%%%%%%%%%%%%%%%%
\section{Supersymmetry analysis}

As shown in \cite{Liu:2004ru}, IIB supergravity admits a consistent
bosonic $S^3\times S^3$ breathing mode reduction of the form
\begin{eqnarray}
ds_{10}^2&=&g_{\mu\nu}(x)dx^\mu dx^\nu+e^{H(x)}(e^{G(x)}
d\Omega_3^2+e^{-G(x)}d\widetilde\Omega_3^2),\nonumber\\
F_{(5)}&=&F_{(2)}\wedge\omega_3+\widetilde F_{(2)}\wedge\widetilde
\omega_3,
\end{eqnarray}
which only retains the relevant fields, namely the metric and self-dual
five-form.  The resulting four-dimensional Lagrangian is given by
\begin{equation}
e^{-1}\mathcal L=e^{3H}[R+\ft{15}2\partial H^2-\ft32\partial G^2
-\ft14e^{-3(H+G)}F_{\mu\nu}^2+12e^{-H}\cosh G].
\label{eq:lag4}
\end{equation}
Although this system is not a consistent supergravity in four
dimensions, it nevertheless has associated with it the reduction
of the IIB supersymmetry variations
\begin{eqnarray}
\delta\psi_\mu&=&[\nabla_\mu+\ft{i}{16}e^{-\fft32(H+G)}F_{\nu\lambda}
\gamma^{\nu\lambda}\gamma_\mu]\epsilon,\nonumber\\
\delta\chi_H&=&[\gamma^\mu\partial_\mu H+e^{-\fft12H}(\eta e^{-\fft12G}
-i\widetilde\eta\gamma_5e^{\fft12G})]\epsilon,\nonumber\\
\delta\chi_G&=&[\gamma^\mu\partial_\mu G-\ft{i}4e^{-\fft32(H+G)}
F_{\mu\nu}\gamma^{\mu\nu}+e^{-\fft12H}(\eta e^{-\fft12G}+i\widetilde\eta
\gamma_5e^{\fft12G})]\epsilon,
\label{eq:susy4}
\end{eqnarray}
where the four-dimensional spinors are Dirac and $\eta=\pm1$,
$\widetilde\eta=\pm1$ are independent signs related to the Killing
spinor orientations on the three-spheres.

The usefulness of this consistent reduction approach is that any
solution to the four-dimensional system (\ref{eq:lag4}) may be
lifted to obtain a IIB solution with the isometry of $S^3\times S^3$.
If, in addition, the solution satisfies the Killing spinor conditions
deduced from (\ref{eq:susy4}), then the uplifted background is guaranteed
to be supersymmetric.  Working with this system is of course equivalent
to working in the original framework of \cite{Lin:2004nb}.

The supersymmetry analysis \cite{Lin:2004nb,Liu:2004ru} proceeds by defining
a complete set of Dirac bilinears
\begin{eqnarray}
&&f_1=\bar{\epsilon}\gamma^5\epsilon,\qquad f_2=i\bar{\epsilon}\epsilon,
\qquad
K^\mu=\bar{\epsilon}\gamma^\mu\epsilon,\qquad
L^\mu=\bar{\epsilon}\gamma^\mu \gamma^5\epsilon,\nonumber\\
&&Y_{\mu\nu}=i\bar{\epsilon}\gamma_{\mu\nu}\gamma^5\epsilon,
\end{eqnarray}
where $\epsilon$ is assumed to be a Killing spinor.
By Fierz rearrangement, we may obtain the algebraic identities 
\begin{equation}
L^{2}=-K^{2}=f_{1}^{2}+f_{2}^{2},\qquad K\cdot L=0. 
\label{eq:fierz}
\end{equation}
As $K^\mu$ turns out to be a Killing vector with non-positive norm,
solutions fall into two classes, depending on whether $K^\mu$ is
timelike or null.  The timelike case was thoroughly investigated in
\cite{Lin:2004nb}.  Hence we only concern ourselves with the null
case in the present analysis.

{}From (\ref{eq:fierz}), it is clear that the null Killing vector
case ($K^2=0$) corresponds to taking $f_1=f_2=0$.  Although it is clear
from above that the vector $L^\mu$ has zero norm, we can further use the
$\delta\chi_H=0$ `differential' identities to deduce that it is in fact
vanishing, {\it i.e.}~$L_\mu=0$.  In this case, the complete set of
differential identities given in Appendix~C of \cite{Liu:2004ru}
collapse into the equivalent set
\begin{eqnarray}
&&\eta K_\mu=*Y_\mu{}^\nu\partial_\nu w_1,\qquad
*Y_\mu{}^\nu\partial_\nu w_2=0,\qquad
2K_{[\mu}\partial_{\nu]}w_1=\eta*Y_{\mu\nu},\nonumber\\
&&\widetilde\eta K_\mu=Y_\mu{}^\nu\partial_\nu w_2,\qquad
~~Y_\mu{}^\nu\partial_\nu w_1=0,\qquad
~2K_{[\mu}\partial_{\nu]}w_2=\widetilde\eta Y_{\mu\nu},\nonumber\\
&&K^\mu\partial_\mu w_1=K^\mu\partial_\mu w_2=0,\qquad
K^\mu F_{\mu\nu}=K^\mu*F_{\mu\nu}=0,\nonumber\\
&&F_{\mu\nu}Y^{\mu\nu}=F_{\mu\nu}*Y^{\mu\nu}=F_\mu{}^\lambda Y_{\nu\lambda}=0,
\nonumber\\
&&\nabla_\mu K_\nu=\nabla_\mu Y_{\nu\lambda}=0.
\label{eq:diffid}
\end{eqnarray}
Here we have defined the scalar combinations
\begin{equation}
w_1\equiv e^{\fft12(H+G)},\qquad w_2\equiv e^{\fft12(H-G)}.
\label{eq:w1w2}
\end{equation}

\subsection{Specialization of the metric}

We now note from (\ref{eq:diffid}) that $K^\mu$ is both null and
covariantly constant.  This allows us to introduce null coordinates
$(u,v)$ such that
\begin{equation}
K^\mu\partial_\mu=\fft\partial{\partial v},\qquad K_\mu dx^\mu=du.
\label{eq:kdef}
\end{equation}
The resulting metric may be written in the form of a pp-wave
\begin{eqnarray}
ds^{2} &=&2\,du\,dv+\mathcal{F}\,du^{2}+\Omega^2(dy_{1}^{2}+dy_{2}^{2})
\nonumber\\
&=&e^{+}e^{-}+e^{i}e^{i},
\label{eq:met1}
\end{eqnarray}
where
\begin{equation}
e^{+}=du,\qquad e^{-}=dv+\ft{1}{2}\mathcal{F}\,du,\qquad
e^{i}=\Omega\,dy_{i}.
\end{equation}
For simplicity, we have chosen the two-dimensional transverse metric to
be conformally flat.  Furthermore, both functions $\mathcal F$ and
$\omega$ are independent of $v$.

We may now use the identities $\widetilde\eta Y=K\wedge dw_2$
and $\eta*Y=K\wedge dw_1$ from (\ref{eq:diffid}) along with (\ref{eq:kdef})
to demonstrate that the two-form $Y$ takes the form
\begin{equation}
Y=(\widetilde\eta\partial_i w_2) du\wedge dy_i=(\eta\epsilon_{ij}
\partial_jw_1) du\wedge dy_i,
\label{eq:ysol}
\end{equation}
where we recall that the signs $\eta$ and $\widetilde\eta$ are of unit
magnitude.  We note that (\ref{eq:ysol}) implies the 
Cauchy-Riemann equations for $(w_1,w_2)$
\begin{equation}
\partial_iw_2=\eta\widetilde\eta\epsilon_{ij}\partial_jw_1
\end{equation}
(which further ensures that they are harmonic in the $y_1$--$y_2$ plane).

Using $\eta K_\mu=*Y_\mu{}^\nu\partial_\nu w_1$ and
$\widetilde{\eta}K_\mu=Y_\mu{}^\nu\partial_\nu w_2$ from (\ref{eq:diffid})
now allows us to obtain the conformal factor
\begin{equation}
\Omega ^{2}=\partial _{i}w_{1}\partial _{i}w_{1}=\partial _{i}w_{2}\partial
_{i}w_{2}.
\end{equation}
The Cauchy-Riemann equations along with the solution for the conformal
factor allow us to make the observation
\begin{equation}
\Omega^2(dy_1^2+dy_2^2)=(dw_1-\partial_uw_1\,du)^2+(dw_2-\partial_uw_2\,du)^2.
\end{equation}
As a result, it becomes natural to perform a ($u$-dependent) coordinate
transformation from $y_i$ to $x_i$ according to
\begin{equation}
x_1=w_1(u,y_1,y_2),\qquad x_2=-\eta\widetilde\eta w_2(u,y_1,y_2),
\end{equation}
where the sign factors in the $x_2$ transformation are chosen for
later convenience.  The result of this transformation is to essentially
make use of $(w_1,w_2)$ as coordinates on the base.  This manipulation
parallels the observation presented in \cite{Gauntlett:2002nw} that a
preferred set of coordinates may be chosen on the base for the null case.

The above coordinate transformation allows us to further specialize the
metric (\ref{eq:met1}) to the particularly simple form
\begin{equation}
ds^2=2\,du\,dv+\mathcal F\,du^2+(dx_1-a_1\,du)^2+(dx_2-a_2\,du)^2.
\end{equation}
Corresponding to this metric, we have
\begin{equation}
w_1=x_1,\qquad w_2=-\eta\widetilde\eta\,x_2,\qquad Y=-\eta\,du\wedge dx^2.
\label{eq:w1w2y}
\end{equation}
A note on the choice of signs is now in order.  We recall from (\ref{eq:w1w2})
that both $w_1$ and $w_2$ are non-negative.  As a result, ($x_1$, $x_2$) are restricted to a single quadrant of the plane.  We may choose both $x_1$ and
$x_2$ to be positive, in which case the signs must obey the condition
$\eta=-\widetilde\eta=\pm1$.  As in \cite{Lin:2004nb}, this two-fold
degeneracy of signs is important in ensuring that the solution is overall
1/2 BPS and not 1/4 BPS.

Before obtaining the two-form field strength $F$, we first
constrain the vector $\vec a(u,x_1,x_2)$ on the base.  Perhaps the most
straightforward way to do so is to enforce the covariant constancy of
$Y$ indicated in (\ref{eq:diffid}).  In particular, $\nabla_uY_{u1}=0$
demonstrates that $\vec a$ is curl-free
\begin{equation}
\epsilon_{ij}\partial_ia_j=0.
\end{equation}
As a result, $\vec a$ may be written as a divergence, $\vec a=\vec\nabla
\varphi$.  In this case, $\vec a$ may be removed by a coordinate
transformation
\begin{equation}
v\to v+\varphi(u,x_1,x_2),
\end{equation}
along with a redefinition
\begin{equation}
\mathcal F\to\mathcal F-2\partial_u\varphi-(\partial_i\varphi)^2.
\end{equation}
The metric is now of a conventional pp-wave form
\begin{equation}
ds^2=2\,du\,dv+\mathcal F\,du^2+dx_1^2+dx_2^2.
\label{eq:metsimple}
\end{equation}
%

%%%%%%%%%%%%%%%%%%%%%%%%%%%%%%%%%%%%%%%%
\section{Completing the solution}

To complete the solution, we now turn to the two-form field strength
$F$.  From $K^\mu F_{\mu\nu}=K^\mu*F_{\mu\nu}=0$ given in
(\ref{eq:diffid}), we see that $F$ only has components along
$du\wedge dx^i$.  However, using $F_\mu{}^\lambda Y_{\nu\lambda}=0$
along with the explicit form of $Y$ in (\ref{eq:w1w2y}), we see that
the $du\wedge dx^2$ component must vanish.  As a result, $F$ may
be written as
\begin{equation}
F=F_{u1}(u,x_1,x_2)\,du\wedge dx^1.
\label{eq:fs1}
\end{equation}

In writing the field strength (\ref{eq:fs1}) and metric (\ref{eq:metsimple}),
we have now exhausted the content of the null case differential identities
(\ref{eq:diffid}).  To guarantee a valid solution, however, we must also
impose the Bianchi identity and equation of motion on $F$
\begin{equation}
dF=0,\qquad d(w_1^{-3}w_2^3*F)=0.
\end{equation}
Using (\ref{eq:fs1}), these are equivalent to
\begin{equation}
\partial_2F_{u1}=0,\qquad\partial_1((x_2/x_1)^3F_{u1})=0.
\end{equation}
As a result, $F$ is given by
\begin{equation}
F=x_1^3f(u)\,du\wedge dx^1.
\end{equation}
For completeness, we note here that the Killing spinor $\epsilon$ is
of the form
\begin{equation}
\epsilon(u)=\exp\left(-\fft{i}4\int^uf(u)\,du\right)\epsilon_0,\qquad
\gamma^+\epsilon_0=0,\qquad\gamma^1\epsilon_0=\epsilon_0,
\end{equation}
where $\epsilon_0$ is a constant spinor.

Finally, as expected for a pp-wave solution, the $uu$ component of the
Einstein equation is undetermined by supersymmetry, and remains to be
imposed.  In the present case, this equation has the form
\begin{equation}
(x_1x_2)^{-3}\partial_i((x_1x_2)^3\partial_i\mathcal F)=-f(u)^2,
\end{equation}
and admits a solution
\begin{equation}
\mathcal F=-\ft1{16}(x_1^2+x_2^2)f(u)^2+\mathcal F_0(u,x_1,x_2),
\label{eq:calfsoln}
\end{equation}
where $\mathcal F_0$ is any solution to the Laplace equation
\begin{equation}
\partial_i((x_1x_2)^3\partial_i\mathcal F_0)=0.
\label{eq:calf0soln}
\end{equation}

The solution is now complete, and is specified by two functions:
an arbitrary function $f(u)$, and a harmonic (in $x_1,x_2$) function
$\mathcal F_0(u,x_1,x_2)$. Although we have worked with an effective
four-dimensional description, (\ref{eq:lag4}), it is instructive to
write down the lifted null solution
\begin{eqnarray}
&&ds^2=2\,du\,dv+\mathcal F\,du^2+[dx_1^2+x_1^2d\Omega_3^2]
+[dx_2^2+x_2^2d\widetilde\Omega_3^2],\nonumber\\
&&F_{5}=x_1^3f(u)\,du\wedge dx_1\wedge\omega_3+x_2^3f(u)\,du\wedge dx_2
\wedge\widetilde\omega_3,
\label{eq:nullsoln}
\end{eqnarray}
where $\mathcal F$ is given by (\ref{eq:calfsoln}) and (\ref{eq:calf0soln}).
It is now clear that we have precisely reproduced the IIB pp-wave
\cite{Blau:2001ne} under the constraint of $S^3\times S^3$ invariance.
The generic pp-wave of this form preserves 16 of the original 32
supersymmetries, although the maximally supersymmetric case may be
obtained by taking $f(u)=\hbox{constant}$ and $\mathcal F_0=0$
\cite{Blau:2001ne}.

We note that the generic IIB pp-wave does not fall into the LLM bubbling
picture of \cite{Lin:2004nb}, although the latter does allow for 1/2 BPS
excitations of plane wave geometries.  The most direct way to see that
the pp-wave excitations are distinct is to perform the change of variables
\begin{equation}
u=t,\qquad v=\hat x_1,\qquad y=x_1x_2,\qquad \hat x_2=\ft12{x_1^2-x_2^2},
\end{equation}
on the ten-dimensional metric (\ref{eq:nullsoln}).  The resulting IIB
geometry then has the LLM form
\begin{equation}
ds^2=-h^{-2}(dt-h^2d\hat x_1)^2+\hat{\mathcal F}dt^2+h^2(d\hat x_1^2
+d\hat x_2^2+dy^2)+y[e^Gd\Omega_3^2+e^{-G}d\widetilde\Omega_3^2],
\end{equation}
where
\begin{equation}
h^{-2}=2y\cosh G,\qquad e^G=\hat x_2/y+\sqrt{(\hat x_2/y)^2+1},
\end{equation}
and
\begin{equation}
\hat{\mathcal F}=(1-\ft1{16}f(t)^2)h^{-2}+\mathcal F_0.
\end{equation}
It is now evident that only when $\hat{\mathcal F}=0$ (corresponding to
the maximally supersymmetric plane wave) does the above solution admit a
strict LLM interpretation.  This furthermore demonstrates that the bubbling
excitations above the plane wave background ({\it i.e.}~the half filled
$\hat x_1$-$\hat x_2$ plane) fall only into the timelike Killing vector
category.  This is the case, even for the constant energy density excitations,
which admit the Killing vector $\partial/\partial\hat x_1$.

%%%%%%%%%%%%%%%%%%%%%%%%%%%%%%%%%%%%%%%%
\section{Conclusions}

We have thus demonstrated that all 1/2 BPS bubbling AdS solutions
(or, more precisely, IIB solutions with $\mathrm{SO}(4)\times
\mathrm{SO}(4)$ isometry) in the null Killing vector case are of the
form of IIB pp-waves.  This null case analysis complements the timelike
case investigated in \cite{Lin:2004nb}, and essentially completes the
study of 1/2 BPS solutions of IIB supergravity with $\mathrm{SO}(4)
\times\mathrm{SO}(4)$ isometry.  Except for the remote possibility that
three-form flux could be turned on without breaking any further
supersymmetry, we have essentially completed the proof that {\it all}
1/2 BPS solutions of IIB supergravity with the given isometry are either
of bubbling AdS \cite{Lin:2004nb} or pp-wave \cite{Blau:2001ne} form.

We note that this analysis of the null Killing vector case of IIB
solutions with $\mathrm{SO}(4)\times\mathrm{SO}(4)$ isometry can
also be extended to the bubbling AdS$_3$ case of minimal
six-dimensional supergravity with $T^2$ isometry
\cite{Martelli:2004xq,Liu:2004hy}.  However, this $\mathcal N=(1,0)$
theory is very much simpler than IIB supergravity, and in fact all
its supersymmetric vacua have already been classified
\cite{Gutowski:2003rg}.  (This is in fact how the bubbling geometries
of \cite{Martelli:2004xq} were obtained.) Thus there is little to be
gained from working out the null analysis, except perhaps a more
explicit means of writing out solutions in a preferred coordinate frame.

Although this exhausts the classification of 1/2 BPS configurations
with $\mathrm{SO}(4)\times\mathrm{SO}(4)$ isometry, the full
classification remains open once the isometry constraint is relaxed.
As mentioned above, the general parallel D3-brane supergravity
solution is an example of a perfectly regular 1/2 BPS solution which
falls outside of the present classification.  It would certainly
be worthwhile, as well as instructive, to obtain a complete construction
of all regular 1/2 BPS solutions of IIB supergravity.  This
construction appears highly non-trivial, as without any underlying
isometry assumption, we have all ten dimensions to consider.  In
contrast, the use of $\mathrm{SO}(4)\times\mathrm{SO}(4)$ isometry
reduced the problem to one of an effective four-dimensional system, and
the application of the invariant tensor method for constructing such
lower dimensional backgrounds is presently very well understood.
(Even though this four-dimensional truncation is not a
consistent supergravity,
all that is needed is a manageable set of Killing spinor conditions,
which this system admits, in order to proceed with the construction.)

While we were initially motivated to seek out novel bubbling
AdS solutions with null isometries, we have ended up focusing on
the classification and potential construction of new 1/2 BPS
solutions regardless of the existence of any gauge theory duals.
This classification program has been an extremely rich one, and
there are of course a much larger variety of possibilities than
just 1/2 BPS structures \cite{Hackett-Jones:2004yi,Gran:2005wn}.
In fact, one of our ultimate aims would be to relax the $\mathrm{SO}(4)
\times\mathrm{SO}(4)$ isometry, and thereby examine backgrounds
with fewer supersymmetries.  We are especially interested in 1/8 BPS
configurations, as they would encompass many important systems such
as $\mathcal N=1$ flux compactifications and models of black hole
microstates, as well as reduced supersymmetry sectors of the gauge/gravity
correspondence.

%%%%%%%%%%%%%%%%%%%%%%%%%%%%%%%%%%%%%%%%
\section*{Acknowledgments}

This work was initiated during a visit to the Khuri lab at the
Rockefeller University.
JTL wishes to acknowledge the hospitality of the KITP where this
work was completed.  This work was supported in part by the US Department
of Energy under grant DE-FG02-95ER40899 and by the National Science
Foundation under grants PHY99-07949 and PHY06-01213.

%%%%%%%%%%%%%%%%%%%%%%%%%%%%%%%%%%%%%%%%

\end{document}